\input{aipcheck}

\documentclass[
    ,final            
  ,numberedheadings 
  ]
  {aipproc}

\usepackage{amsmath}

\layoutstyle{6x9}

\begin{document}

\title{Determining chiral couplings at NLO}

\thanks{Talk given at the International Workshop on Quantum Chromodynamics QCD@work 2007, 16th -- 20th June (2007), Martina Franca (Italy).}


\author{Ignasi Rosell}{
  address={Departamento de Ciencias F\'\i sicas, Matem\'aticas y de la Computaci\'on, Universidad CEU Cardenal Herrera, San Bartolom\'e 55, E-46115 Alfara del Patriarca, Val\`encia, Spain}
}

\begin{abstract}
We present a general method that allows to estimate the low-energy constants of Chiral Perturbation Theory up to next-to-leading corrections in the $1/N_C$ expansion, that is, keeping full control of the renormalization scale dependence. As a first step we have determined $L_8$ and $C_{38}$, the couplings related to the difference of the two-point correlation functions of two scalar and pseudoscalar currents, $L_8^r(\mu_0)=\left(0.6\pm 0.4 \right) \cdot 10^{-3}$ and $C_{38}^r(\mu_0)=\left(2\pm6\right)\cdot 10^{-6}$, with $\mu_0=0.77$~GeV. As in many effective approaches, one of the main ingredients of this method is the matching procedure: some comments related to this topic are presented here.
 \end{abstract}
\maketitle

\section{Introduction}

Chiral Perturbation Theory ($\chi$PT) is the low-energy effective field theory of QCD~\cite{ChPT1,ChPT1bis}. This framework is based on a perturbative expansion in the momenta and masses of the pseudo-Goldstone bosons, the well-known chiral expansion, and it has proved to be a controlled and useful scheme. 

As in many effective approaches only a finite number of terms in the lagrangian is needed to work up to a fixed chiral order. Therefore, the more precision is desired the more couplings are needed. At leading order (LO), and apart from masses, only two couplings appear ($F$ and $B_0$, related to the decay constant of the pion and the quark condensate respectively). However, at next-to-leading order (NLO) 7 and 10 couplings are required for the two and three flavor case respectively. Finally, at next-to-next-to-leading order (NNLO) 53 and 90 couplings are contained in the lagrangian depending again on the considered number of flavors~\cite{ChPT2}. These are the familiar chiral low-energy constants (LECs). Although at $\mathcal{O}(p^4)$ (NLO) one can extract the couplings from the phenomenology, the large number of unknown LECs appearing at $\mathcal{O}(p^6)$ (NNLO) is one of the major problems for the two-loop order calculations~\cite{ChPT3}. Therefore it is clear that a very appealing task is the theoretical estimation of these parameters.

Taking into account that the couplings of any effective theory have information from the degrees of freedom that have been integrated out, the use of Resonance Chiral Theory (R$\chi$T) has been proposed as a way to get information of the chiral couplings~\cite{RChT,RChTb}. Actually, the main aim of R$\chi$T in the first articles was to get the leading resonance contributions to the LECs of the $\mathcal{O}(p^4)$ $\chi$PT lagrangian~\cite{RChT}, and the result was very encouraging, since the integration of the lightest resonances seems to saturate the phenomenological values of these couplings. An extension of this result up to $\mathcal{O}(p^6)$ has been analyzed in Ref.~\cite{RChTb}. This outcome has been rigorously understood and analyzed in the context of the $1/N_C$ expansion. In fact, one of the main ingredients to deal with resonance chiral lagrangians is the $1/N_C$ expansion.

Remembering the difficulties of a formal Effective Field Theory method in the resonance region, our effective approach is based on the phenomenological lagrangians'~ideas of Ref.~\cite{ChPT1}, {\it i.e.}, considering the most general possible lagrangian, including all terms consistent with assumed symmetry principles. This approach is realized by making use of the $1/N_C$ expansion and the short-distance constraints from QCD. The limit of a large number of colors turns out to be a very helpful tool in order to understand many features of QCD and becomes an alternative power counting. Assuming confinement, the $N_C \rightarrow \infty$ limit guarantees that meson dynamics are described by tree-level interactions of an effective local lagrangian including only meson degrees of freedom, higher corrections in $1/N_C$ being obtained by loop corrections~\cite{NC}. Furthermore, the short-distance properties of QCD constraint strongly the lagrangian and provide relations between the couplings.

It is important to highlight that the only model dependence of our method is the cut of the tower of resonances, that is, although large-$N_C$ QCD needs an infinite number of resonances to recover the usual QCD results, we approximate the result by truncating the tower to the lowest states. It seems a reasonable assumption bearing in mind that heavier contributions are suppressed by their masses. Moreover, the phenomenology supports this approximation.

Note the appealing application of R$\chi$T to estimate chiral couplings. Although the number of couplings is much higher in R$\chi$T than in $\chi$PT, the underlying information from QCD reduces the number of unknown coupling and the low-energy limit gives a prediction of the LECs in terms of a small number of parameters, that can be obtained phenomenologycally.

Since chiral loops are of NLO in the $1/N_C$ expansion, the LO estimation of the LECs is not able to control the renormalization-scale dependence. Despite the fact that this is not so important in the vector sector, it produces remarkable uncertainties for the scalar one. This is one of the main motivations to consider quantum loops within resonance chiral lagrangians. In addition take note of the necessity of improving the hadronic contributions to distinguish new physics effects from Standard Model results.

\section{Short-distance constraints}

One of the main features of R$\chi$T, and what makes so interesting our effective procedure, is the use of the short-distance information from QCD. Actually, as previously pointed out, there are too many unknown parameters without examining the high-energy properties of the underlying strong dynamics. Basically there are two sources of information:
\begin{enumerate}
 \item One considers the Green functions of QCD currents obtained in the Operator Product Expansion (OPE). In other words, the effective and OPE results are matched. The resonance result is expanded at high energies and the first coefficients of the expansion are compared with the OPE ones. Note that the use of a matching point is meaningless in this case taking into account the cut of the tower of resonances. Speaking about matching `at large energies' is an abuse of notation, since the OPE usually works at $2$~GeV and one is thinking about this scale of energies in the matching procedure. 
\item The form factors are the other way to constraint the lagrangian, {\it i.e.}, to demand that two-body form factors of hadronic currents vanish at high energies. This assumption is clear in the case of pseudo-Goldstone bosons (the pion multiplet) and photons, since it has been observed experimentally. The case of resonances as asymptotic states is an {\it a priori} open question. We think that the consideration of vanishing form factors at large momentum transfer is a reasonable conjecture. This supposition was motivated in our case by the calculation of the vector form factor of the pion at one-loop level~\cite{VFF_NLO}, where it seemed to be a necessary requirement to recover a well-behaved result. Within parton dynamics this behavior has been understood, the so-called Brodsky-Lepage rules~\cite{Brodsky-Lepage}. 

It is better understood in the context of two-point correlation functions of two QCD currents at NLO~\cite{formfactors,formfactors2}, whose spectral functions are related to the form factors by using the optical theorem. It is much clearer in the case of the vector or axial-vector currents, since at high energies the spectral function tends to a constant. Taking into consideration that the spectral function is a sum of positive contributions corresponding to the different intermediate states and there is an infinite number of possible states, the absorptive contribution of a given state should vanish at infinite momentum transfer. In the scalar and pseudoscalar case the spectral function grows as $t$ at large energies. Following the previous argument, it would mean that the absorptive contribution of each state should behave as a constant at short distance. However, the $SS-PP$ sum rules, the Brodsky-Lepage ideas and the $1/t$ behavior of each one-particle intermediate cut seem to indicate that the vanishing assumption is reasonable. In the spirit of correlators at subleading order, short-distance constraints coming from form factors with resonances are only mandatory for one-loop calculations.
\end{enumerate}
 
Since there is an infinite number of Green functions, it is obviously not possible to satisfy all matching conditions with a finite number of resonances. As pointed out previously, the results obtained with this procedure are model dependent only in this sense: the tower of resonances has been cut. Take note that in order to recover the large-$N_C$ behavior of QCD an infinite number of states is needed. Therefore, it is not surprising to find some conflicts between different constraints. The general procedure is including more resonances if one incompatibility appears, following the Minimal Hadronic Approximation ideas of Refs.~\cite{MHA}.

\section{Saturation at NLO}

The aim of this work is to estimate the chiral couplings by expanding the results of R$\chi$T at low energies. In the case of the $\mathcal{O}(p^4)$ couplings, this expansion reads:
\begin{eqnarray}
 L_i(\mu)&=&L_i^R(\mu) + \widetilde{L}_i(\mu)\,,
\end{eqnarray}
being $L_i$ the $\chi$PT coupling, $\widetilde{L}_i$ the R$\chi$T ones and $L_i^R$ the contributions coming from the low-energy expansion of the resonance contributions. The saturation means that $\widetilde{L}_i(\mu)=0$, that is, one does not need to consider the equivalent coupling in the theory with resonances. See that this fact is very interesting to avoid new unknown parameters. 

At LO in the $1/N_C$ the asymptotic behavior of QCD require that $\widetilde{L}_i=0$ and it is found that~\cite{RChT}:
\begin{align}
&L_1=\frac{G_V^2}{8M_V^2}\,,\quad  
L_2=\frac{G_V^2}{4M_V^2}\,, \quad
L_3=-\frac{3G_V^2}{4M_V^2}\,+\,\frac{c_d^2}{2M_S^2}\, ,\nonumber \\
&L_5=\frac{c_dc_m}{M_S^2}\,,\quad
L_8=\frac{c_m^2}{2M_S^2}\,-\,\frac{d_m^2}{2M_P^2} \,, \quad
L_9=\frac{F_VG_V}{2M_V^2}\,, \nonumber \\
&L_{10}=-\frac{F_V^2}{4M_V^2}\,+\,\frac{F_A^2}{4M_A^2}\,,\qquad
L_4=L_6=L_7=0\,, \label{rescont}
\end{align}

The present project basically consists on doing the same at NLO. One needs to know if the saturation is valid at subleading order. In Ref.~\cite{vanishing} it is proved in the case of $\widetilde{L}_4$, $\widetilde{L}_5$, $\widetilde{L}_8$ and $\widetilde{L}_6 + \widetilde{L}_7$, the couplings related to the scalar and pseudoscalar sector. It is tempting to extrapolate these results to other couplings, where there are two troubles: the introduction of vector and axial-vector mesons and the study of the scattering amplitudes~\cite{vanishing2}.  

\section{$L_8^r(\mu)$ and $C_{38}^r(\mu)$}

As a first step in the determination of chiral couplings at NLO we present a one-loop estimation of the couplings $L_8^r(\mu)$ and $C_{38}^r(\mu)$~\cite{L8}, the couplings related to $\Pi(t)\equiv \Pi_S(t) - \Pi_P(t)$, the difference of the two-point correlation functions of two scalar and pseudoscalar currents respectively, which is identically zero in perturbative QCD. At very low energies, $\Pi(t)$ is calculated within $\chi$PT and reads~\cite{ChPT1bis,ChPT2}
\begin{eqnarray}
  \Pi(t) &=&  B_0^2 \, \left\{ \frac{2 F^2}{t} \, +\, 32 L_8^r(\mu)
\, +\, \frac{\Gamma_8}{\pi^2} \left( 1-\log \frac{-t}{\mu^2} \right)\right. \nonumber \\ &&
\quad +\, 
\frac{t}{F^2} \bigg[ 32\,C_{38}^r (\mu)  -\frac{\Gamma_{38}^{(L)}}{\pi^2}  \left( 1-\log \frac{-t}{\mu^2} \right) +\mathcal{O}\left(N_C^0\right) \bigg]\,+\,
 \mathcal{O}\left(t^2\right)\Bigg\}\, ,\label{eq:Pi_chpt}
\end{eqnarray}
with $\Gamma_8 = 5/48$ [$3/16$] and $\Gamma_{38}^{(L)}=-5L_5^r/6$ [$-3L_5^r/2$] in the $SU(3)_L\otimes SU(3)_R$
[$U(3)_L\otimes U(3)_R$] effective theory~\cite{ChPT1bis,ChPT2}. Note that $F^2$, $L_8^r$ and $C_{38}^r/F^2$ are $\mathcal{O}(N_C)$, being $\Gamma_8$ and $\Gamma_{38}^{(L)}/F^2$ $\mathcal{O}(1)$, so we are discarding the NNLO contributions, as it should.

At higher energies, $\Pi(t)$ is obtained in the context of R$\chi$T,
\begin{eqnarray}
 \Pi(t)&=& 2\, B_0^2\,\left\{
\frac{8\, c_{m}^{r\, 2}}{M_{S}^{r\, 2} - t}\, -\,
\frac{8\, d_{m}^{r\, 2}}{M_{P}^{r\,  2} - t} \,+\, \frac{F^2}{t}
\,+\, \Delta\Pi(t)|_\rho\right\}\, ,\label{result}
\end{eqnarray}
where $c_m^r$, $d_m^r$, $M_S^r$ and $M_P^r$ are resonance couplings at NLO and $\Delta \Pi (t)|_\rho$ can be expressed in terms of resonance masses once the short-distance constraints have been implemented~\cite{L8}.

The short-distance behavior of the one-loop contribution is
\begin{eqnarray}
 \Delta  \Pi(t)|_\rho &=&  \frac{F^2}{t}\,
\delta_{_{\rm NLO}}^{(1)} \, +\, \frac{F^2 M_S^2}{t^2} \,\left( \delta_{_{\rm
NLO}}^{(2)}\, +\, \widetilde{\delta}_{_{\rm NLO}}^{(2)} \,
\log\frac{-t}{M_S^2}\right)\, +\, \mathcal{O}\left(\frac{1}{t^3}\right)\, ,\label{expansion}
\end{eqnarray}
where the corrections $\delta_{\mathrm{NLO}}^{(m)}$ and $\widetilde{\delta}_{\mathrm{NLO}}^{(m)}$ are functions of resonance masses~\cite{L8}. Taking into account that the logarithmic term in Eq.~(\ref{expansion}) should vanish, a relation between resonance masses is got and considering the short-distance behavior in $\Pi(t)$ one finds
\begin{eqnarray}
F^2\, (1\,+\,\delta_{_{\rm NLO}}^{(1)})\, - \, 8\, c_m^{r\, 2} \, +\, 8 \,
d_m^{r\, 2}\, = \, 0 \, ,  \nonumber
\\
F^2 \,M_S^2 \,\delta_{_{\rm NLO}}^{(2)}\, - \, 8\, c_m^{r\, 2}\, M_S^{r\, 2} \,
+\, 8 \, d_m^{r\, 2}\, M_P^{r\, 2} \, = \, -8\, \widetilde{\delta} \, , \label{eq:NLO_rel}
\end{eqnarray}
with $\widetilde{\delta}$ a small correction. Eq.~(\ref{eq:NLO_rel}) fixes $c_m^r$ and $d_m^r$,
\begin{eqnarray}
c_m^{r\,\,2}&=& \frac{F^2}{8} \frac{M_P^{r\,\,2}}{M_P^{r\,\,2}-M_S^{r\,\,2}} \left(1+\delta_{_{\rm NLO}}^{(1)}-\frac{M_S^2}{M_P^2}\delta_{_{\rm NLO}}^{(2)}-\frac{8}{M_P^2 F^2} \widetilde{\delta} \right) \, ,\nonumber \\
d_m^{r\,\,2}&=& \frac{F^2}{8} \frac{M_S^{r\,\,2}}{M_P^{r\,\,2}-M_S^{r\,\,2}} \left(1+\delta_{_{\rm NLO}}^{(1)}-\delta_{_{\rm NLO}}^{(2)} -\frac{8}{M_S^2 F^2} \widetilde{\delta}\right) \, .\label{dmr}
\end{eqnarray}

The comparison of Eq.~(\ref{eq:Pi_chpt}) and the low-momentum expansion of Eq.~(\ref{result}) allows to estimate $L_8^r$ and $C_{38}^r$. After integrating out the singlet $\eta_1$ field, one finally obtains
\begin{eqnarray}
L_8^r(\mu_0) &= & (0.6\,\pm\, 0.4)\cdot 10^{-3}\, , \\
 C_{38}^{r}(\mu_0) &=&\left( 2\,\pm\,6 \right)\cdot 10^{-6}
 \, .
\end{eqnarray}
being $\mu_0$ the usual $\chi$PT scale, $\mu_0=0.77$ GeV.

$L_8^r(\mu_0)$ can be compared with the value $L_8^r(\mu_0)=(0.9\pm 0.3)\cdot
10^{-3}$~\cite{RChT}, usually adopted in $\mathcal{O}(p^4)$ phenomenological analyses, or $L_8^r(\mu_0)=(0.62\pm 0.20)\cdot 10^{-3}$, got from  the $\mathcal{O}(p^6)$ fit of Ref.~\cite{ChPT2loopsdbis}. It is not surprising to be closer to the $\mathcal{O}(p^6)$ fit, since the leading and subleading contributions in the $1/N_C$ expansion have been included in our procedure, as occurred in Ref.~\cite{ChPT2loopsdbis} and not in Ref.~\cite{RChT}.

\section{Conclusions}

The theoretical estimation of the chiral couplings is a very interesting task. In the light of the effective field theory ideas, the use of phenomenological lagrangians with resonances as active degrees of freedom seems to be the appropriate way to do it. Resonance Chiral Theory provides a correct framework to incorporate the resonances and the success to determine the $\mathcal{O}(p^4)$ chiral LECs is striking. The $1/N_C$ expansion and the short-distance constraints are fundamental to understand and arrive to these results.

We present a general procedure to investigate chiral LECs at subleading order in the $1/N_C$ expansion. It is important to stress the importance of considering form factors with resonances as asymptotic states in order to find relations between resonance parameters. 

As a first advance we have estimated $L_8^r$ and $C_{38}^r$~\cite{L8}, the couplings appearing in the difference of the two-point correlation functions of two scalar and pseudoscalar currents respectively. More work is in progress~\cite{formfactors2}.

\begin{theacknowledgments}
  We wish to thank the organizers of the International Workshop on Quantum Chromodynamics QCD at Work 2007 for the useful and pleasant congress and A.~Pich an J.J.~Sanz-Cillero for their interesting suggestions and comments. This work has been supported in part by the EU MRTN-CT-2006-035482 (FLAVIAnet), by MEC (Spain) under grant FPA 2004-00996 and by Generalitat Valenciana under grant GVACOMP2007-156.
\end{theacknowledgments}


\begin{thebibliography}{90}

\bibitem{ChPT1}
  S.~Weinberg,
  Physica A {\bf 96} (1979) 327.

\bibitem{ChPT1bis}
  J.~Gasser and H.~Leutwyler,
  Annals Phys.\  {\bf 158} (1984) 142;
  Nucl.\ Phys.\  B {\bf 250} (1985) 465;
  Nucl.\ Phys.\  B {\bf 250} (1985) 517;
  Nucl.\ Phys.\  B {\bf 250} (1985) 539.

\bibitem{ChPT2}
  J.~Bijnens, G.~Colangelo and G.~Ecker,
  JHEP {\bf 9902} (1999) 020
  [arXiv:hep-ph/9902437];
  Annals Phys.\  {\bf 280} (2000) 100
  [arXiv:hep-ph/9907333].

\bibitem{ChPT3}
  J.~Bijnens,
  Prog.\ Part.\ Nucl.\ Phys.\  {\bf 58} (2007) 521
  [arXiv:hep-ph/0604043].

\bibitem{RChT}
  G.~Ecker, J.~Gasser, A.~Pich and E.~de Rafael,
  Nucl.\ Phys.\  B {\bf 321} (1989) 311;\\
  G.~Ecker, J.~Gasser, H.~Leutwyler, A.~Pich and E.~de Rafael,
  Phys.\ Lett.\  B {\bf 223} (1989) 425.

\bibitem{RChTb}
  V.~Cirigliano, G.~Ecker, M.~Eidem\"uller, R.~Kaiser, A.~Pich and J.~Portol\'es,
  Nucl.\ Phys.\  B {\bf 753} (2006) 139
  [arXiv:hep-ph/0603205].

\bibitem{NC}
  G.~'t Hooft,
  Nucl.\ Phys.\  B {\bf 72} (1974) 461;
  Nucl.\ Phys.\  B {\bf 75} (1974) 461;\\
  E.~Witten,
  Nucl.\ Phys.\  B {\bf 160} (1979) 57.

\bibitem{VFF_NLO}
  I.~Rosell, J.~J.~Sanz-Cillero and A.~Pich,
  JHEP {\bf 0408}, 042 (2004)
  [arXiv:hep-ph/0407240].

\bibitem{Brodsky-Lepage}
  G.~P.~Lepage and S.~J.~Brodsky,
  Phys.\ Lett.\  B {\bf 87} (1979) 359;
  Phys.\ Rev.\  D {\bf 22} (1980) 2157.

\bibitem{formfactors}
  I.~Rosell, Ph.D. thesis, Universitat de Val\`encia (2007),
  arXiv:hep-ph/0701248.

\bibitem{formfactors2}
 I.~Rosell, J.~J.~Sanz-Cillero and A.~Pich, work in preparation.

\bibitem{MHA}
  S.~Peris, M.~Perrottet and E.~de Rafael,
  JHEP {\bf 9805} (1998) 011
  [arXiv:hep-ph/9805442]; \\
  M.~Knecht and E.~de Rafael,
  Phys.\ Lett.\ B {\bf 424} (1998) 335
  [arXiv:hep-ph/9712457]; \\
  M.~Knecht, S.~Peris and E.~de Rafael,
  Phys.\ Lett.\ B {\bf 443} (1998) 255
  [arXiv:hep-ph/9809594]; \\
  M.~Golterman, S.~Peris, B.~Phily and E.~De Rafael,
  JHEP {\bf 0201} (2002) 024
  [arXiv:hep-ph/0112042]; \\
  M.~Golterman and S.~Peris,
  JHEP {\bf 0101} (2001) 028
  [arXiv:hep-ph/0101098].

\bibitem{vanishing}
  J.~Portol\'es, I.~Rosell and P.~Ruiz-Femen\'\i a,
  Phys.\ Rev.\  D {\bf 75}, 114011 (2007)
  [arXiv:hep-ph/0611375].

\bibitem{vanishing2}
  I.~Rosell, P.~Ruiz-Femen\'\i a and J.~J.~Sanz-Cillero, work in preparation.

\bibitem{L8}
  I.~Rosell, J.~J.~Sanz-Cillero and A.~Pich,
  JHEP {\bf 0701} (2007) 039
  [arXiv:hep-ph/0610290].

\bibitem{ChPT2loopsdbis}
  G.~Amor\'os, J.~Bijnens and P.~Talavera, 
  Nucl.\ Phys.\ B {\bf 602} (2001) 87
  [arXiv:hep-ph/0101127].

\end{thebibliography}
\end{document}